\documentclass{article}

\usepackage[sort]{natbib}
\usepackage{epsfig,amsmath,amssymb}

\newcommand{\ignore}[1]{}

\renewcommand{\deg}[1]{d_{#1}}
\newcommand{\matr}[1]{\mathbf #1}
\newcommand{\A}{\matr{A}}
\newcommand{\I}{\matr{I}}
\newcommand{\D}{\matr{D}}

\renewcommand{\P}{\matr{P}}
\newcommand{\Q}{\matr{Q}}
\newcommand{\M}{\matr{M}}
\renewcommand{\L}{\matr{L}}
\newcommand{\U}{\matr{U}}

\newcommand{\CP}{{\cal P}}
                      \newcommand{\CQ}{\cal{Q}}
\newcommand{\CL}{\cal{L}}
\newcommand{\HP}{\hat{\P}}
\newcommand{\vect}[1]{\mathbf{#1}}
\renewcommand{\u}{\vect{u}}
\newcommand{\vv}{\vect{v}}
\newcommand{\m}{\vect{m}}
\renewcommand{\d}{\vect{d}}

\newcommand{\inv}[1]{{#1}^{-1}}
\newcommand{\transpose}[1]{{#1}^{\scriptscriptstyle\mathrm T}}
\renewcommand{\cal}[1]{\mathcal{#1}}
\renewcommand{\sqrt}[1]{{#1}^{\frac{1}{2}}}
\newcommand{\invsqrt}[1]{{#1}^{-\frac{1}{2}}}
\newcommand{\superscript}[1]{\ensuremath{^\textrm{#1}}}
\renewcommand{\th}{\superscript{th}}
\newcommand{\eig}{\lambda}
\newcommand{\eigval}[2]{\eig^{(#1)}_{#2}}

\newcommand{\leig}{\mu}
\newcommand{\leigval}[2]{\leig^{(#1)}_{#2}}

\newcommand{\spec}[1]{\textrm{Spec}(#1)}

\newcommand{\argmin}{\textrm{argmin}}
\newcommand{\diag}{\textrm{diag}}
\newcommand{\constant}{c}

\newcommand{\Chat}{\hat{C}}

\newcommand{\one}{\mathbf{1}}
\newcommand{\RR}{\mathbb{R}}
\newcommand{\Sbar}{\bar{S}}
\newcommand{\Scut}{(S,\Sbar)}
\newcommand{\vf}{\vv^f}
\newcommand{\Vol}{\mbox{Vol}}

\title{Locally computable approximations for spectral clustering and
  absorption times of random walks}

\author{Pekka Orponen$^\dagger$, Satu Elisa Schaeffer$^{\ddagger, 1}$, and Vanesa Avalos-Gayt{\'{a}}n$^\ddagger$ \\
$^\dagger$ {\small Department of Information and Computer Science,} \\
{\small Helsinki University of Technology TKK,}\\
{\small FI-02015 TKK Espoo, Finland} \\
{\small \tt pekka.orponen@tkk.fi} \\
$^\ddagger$ {\small School of Mechanical and Electrical Engineering,} \\
{\small Universidad Aut\'{o}noma de Nuevo Le\'{o}n (UANL),} \\
{\small Ciudad Universitaria, San Nicol\'{a}s de los Garza, NL 66450, Mexico,} \\
{\small \tt $\{$elisa,vanesa$\}$@yalma.fime.uanl.mx} \\
$^1$ {\small Corresponding author, tel.\ +52 81 1340 4000, fax +52 81 1052 3321}}

\date{\today}

\begin{document}

\maketitle

\begin{abstract}
  We address the problem of determining a natural local neighbourhood
  or ``cluster'' associated to a given seed vertex in an undirected
  graph.  We formulate the task in terms of absorption times of random
  walks from other vertices to the vertex of interest, and observe
  that these times are well approximated by the components of the
  principal eigenvector of the corresponding fundamental matrix of the
  graph's adjacency matrix. We further present a locally computable
  gradient-descent method to estimate this Dirichlet-Fiedler vector,
  based on minimising the respective Rayleigh quotient. Experimental
  evaluation shows that the approximations behave well and yield
  well-defined local clusters.
\end{abstract}



\pagestyle{plain} 

\section{Introduction and motivation}

\subsection{Nonuniform networks}

The field of natural-network study became popular when Watts and
Strogatz \cite{WaSt98} published their observations on the short
average path length and the high clustering coefficient of many
natural graphs, followed by the observations of scale-free
distributions \cite{BaAl99,FFFa99} in the degrees and other structural
properties of such networks. As a consequence of the resulting wide
interest in the properties of natural networks, there now exist
numerous models to meet the observations made on natural networks
\cite{DoMe03,Newm03,Virt03}. 

\subsection{Graph clustering}

One of the properties of interest in the field of natural graphs is
the presence of \emph{clusters} or \emph{communities} \cite{NeGi03},
that is, the existence of dense induced subgraphs that have relatively
few connections outside compared to the internal density
\cite{Klei01}.

\begin{figure}
\centerline{\includegraphics[width=50mm]{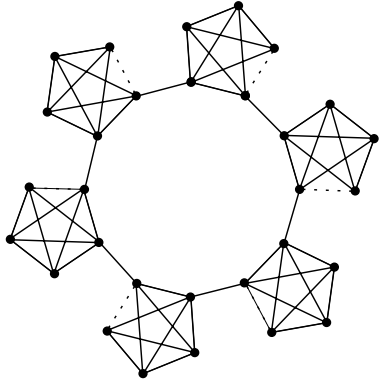}}
  \caption{A \emph{caveman graph} \cite{Watt99} composed of six
    near-cliques of five vertices each that have been connected into a
    circulant graph by ``opening'' one edge from each clique (the
    removed edge is shown with a dotted line).}
\label{fig:caveman}
\end{figure}

\begin{figure}
\centerline{\includegraphics{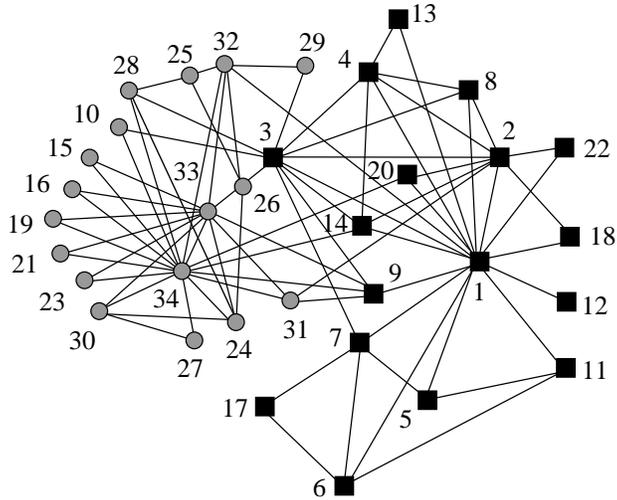}}
\caption{The karate club social network studied by Zachary
  \cite{Zach77}. The two groups into which the club split are
  indicated by the shape with which the vertices are drawn: the
  squares later formed their own club, and the circles formed another
  club.}
\label{fig:karate}
\end{figure}

\emph{Graph clustering} is the task of grouping the vertices of the
graph into clusters taking into consideration the edge structure of
the graph in such a way that there should be many edges \emph{within}
each cluster and relatively few \emph{between} the clusters. For an
artificial example, see Figure \ref{fig:caveman} that illustrates a
small graph with a clear six-cluster structure. Another classic
example is a small real-world social network studied by Zachary
\cite{Zach77} and often referred to in graph clustering papers
\cite{WuHu04,OrSc05,Newm03}. It is a social network of a small karate
club that was just about to split into two (see Figure
\ref{fig:karate}), making it an ideal case for two-classification
algorithms. For a survey on graph-clustering algorithms, see
\cite{Scha07}.

\subsection{Local clustering}

In \emph{local clustering}, the goal is to find the cluster of a given
\emph{seed vertex} $s \in V$. Hence, essentially, it is the task of
finding a \emph{bipartition} of the graph $G$ into two vertex sets $S$
and $V \setminus S$ such that $s \in S$ and $S$ makes a good cluster
in some predefined sense. Common cluster quality criteria include {\em
  cut capacity} and related measures such as \emph{conductance}
\cite{SiSc06} or density-based measures \cite{Scha05}. Also methods
motivated by electric networks have been proposed for global and local
clustering alike \cite{WuHu04,OrSc05,NeGi04}.

\subsection{Spectra of graphs}

Let $G = (V, E)$ be an unweighted undirected connected graph with at
least two vertices.  For simplicity, we focus on unweighted graphs,
although much of what follows can easily be generalised to the
weighted case.  Denote the order of $G$, i.e.\ its number of vertices,
by $n$ and identify each vertex $v$ with a label in $\{1, 2, \ldots,
n\}$.  Denote the seed vertex by $s$.  The \emph{adjacency matrix} of
$G$ is the binary matrix $\A$, where $a_{ij} = 1$ if edge $\{i,j\}$ is
in $E$, and otherwise $a_{ij} = 0$.  

For a weighted graph, one would consider instead the analogous
\emph{edge-weight matrix}. Note also that for multigraphs, edge
multiplicities can in the present context be considered simply as
integer weights. For an undirected graph, the adjacency (resp.\ edge
weight) matrix is symmetric, whereas directed graphs pose further
complications in the algebraic manipulation --- we refer the reader to
the textbook and other works of Chung
\cite{Chun97,dirlocal,AnCh07,heatkernel} for properties and local
clustering of directed graphs.

The \emph{degree} $\deg{v}$ of a vertex $v$ is the number (resp.\ total
weight) of its incident edges; thus the components of the \emph{degree
vector} $\d$ of $G$ are the row sums of $\A$.  Denote by $\D$ the
diagonal $n\times n$ matrix formed by setting the diagonal elements
to $d_{ii} = \deg{i}$ and all other elements to zero.

Let $\I$ be the $n\times n$ unit matrix. The \emph{Laplacian} matrix
of $G$ is $\L = \D - \A$ and the \emph{normalised Laplacian} matrix of
$G$ is $\CL = \invsqrt{\D} \L \invsqrt{\D} = \I - \invsqrt{\D} \A
\invsqrt{\D}$.  Since both $\L$ and $\CL$ are symmetric, all their
eigenvalues are real.  It turns out that $\CL$ is in some respects a
more natural object of study than $\L$, and we shall mostly focus on
that. It is easy to see that zero is an eigenvalue of both $\L$ and
$\CL$, and for $\CL$ it can be shown that all the other $n-1$
eigenvalues (counting multiplicities) lie in the interval $[0,2]$.
Denote these in increasing order as $0 = \leig_0 \leq \leig_1 \leq
\dots \leq \leig_{n-1} \leq 2$, and let $\u_i$ be some right
eigenvector associated to $\leig_i$. We may assume that the distinct
eigenvectors $\u_i$ are orthogonal to each other.  For more
information on the spectral and algebraic properties of graphs, see
e.g.\ the excellent monographs of Biggs \cite{Bigg94} and Chung
\cite{Chun97}.

\subsection{Random walks}
\label{sec:random walks}

The \emph{simple random walk} on a graph $G$ is a Markov chain where
each vertex $v \in V$ corresponds to a state and the transition
probability from state $i$ to state $j$ is $p_{ij} = \inv{\deg{i}}$ if
$\{i,j\} \in E$ and zero otherwise. For a weighted graph, $p_{ij}$ is
the ratio of the weight of edge $\{i,j\}$ to the total weight of edges
incident to $i$.

Denote the transition probability matrix of this Markov chain by $\P =
\D^{-1} \A$.  Note that even for undirected graphs, $\P$ is not
in general symmetric. However, it is similar to the matrix
\begin{equation}
\CP = \sqrt{\D} \P \invsqrt{\D} = \invsqrt{\D} \A \invsqrt{\D}
\label{similar}
\end{equation}
which \emph{is} symmetric because $\A$ is the adjacency matrix of an
undirected graph. Thus, $\P$ and $\CP$ have the same spectrum of
eigenvalues, which are all real. Moreover,
\begin{equation}
\begin{array}{rcl}
\displaystyle
\CL
& = & \invsqrt{\D} \L \invsqrt{\D}
  =   \invsqrt{\D} (\D - \A) \invsqrt{\D} \\
\displaystyle
& = & \invsqrt{\D} (\D - \D\P) \invsqrt{\D}
  =   \I - \sqrt{\D}\P\invsqrt{\D} \\
\displaystyle
& = & \I - \CP.
\end{array}
\label{eq:lapprob}
\end{equation}
Consequently, $\eig$ is an eigenvalue of the normalised transition matrix $\CP$
if and only if $\leig = 1 - \eig$ is an eigenvalue of the normalised
Laplacian matrix $\CL$.  Thus, $\P$, $\CP$ and $\CL$ have the following
correspondence: $\vv$ is a right eigenvector associated to eigenvalue $\eig$ in
$\P$ if and only if $\u = \sqrt{\D}\vv$ is a right eigenvector associated
to the same eigenvalue in $\CP$, and to eigenvalue $1 - \eig$ in
$\CL$.

Since in the case of Markov chains, \emph{left} eigenvectors are also
of interest, let us note in passing that the analogous correspondence
holds between each left eigenvector $\pi$ of $\P$ and left eigenvector
$\rho = \pi\invsqrt{\D}$ of $\CP$ or $\CL$.

Denote the eigenvalues of $\P$ in decreasing order as $\eigval{\P}{0}
\geq \eigval{\P}{1} \geq \dots \geq \eigval{\P}{n-1}$.  Since $\P$ is
a stochastic matrix, it always has eigenvalue $\eigval{\P}{0} = 1$,
corresponding to the smallest Laplacian eigenvalue $\leigval{\CL}{0} =
0$. All the other eigenvalues of $\P$ satisfy $|\eigval{\P}{i}| \leq
1$. If moreover $G$ is connected and not bipartite, the Markov chain
determined by $\P$ is ergodic, in which case $|\eigval{\P}{i}| < 1$
for all $i \geq 1$. Without much loss of generality, we shall
assume this condition, and moreover that all the eigenvalues $\eigval{\P}{i}$
are nonnegative. Both of these conditions can be enforced by
considering, if necessary, instead of $\P$ the ``lazy random walk''
with transition matrix
\begin{equation}
\P' = \frac{1}{2}(\I + \P).
\end{equation}
For a connected graph $G$ this chain is ergodic, and has
nonnegative eigenvalues
\begin{equation}
\eigval{\P'}{i} = \frac{1}{2}(1+\eigval{\P}{i}),
\end{equation}
with the same eigenvectors as $\P$.

Let us then consider a transition matrix $\HP$ obtained from $\P$
by making a given state, or vertex $s$ \emph{absorbing}.
Thus, $\HP$ is otherwise equal to $\P$, but all
$\hat{p}_{si} = 0$ except for $\hat{p}_{ss} = 1$.
We shall henceforth assume, for simplicity of notation, that
$s = n$, so that in particular $\HP$ has the block structure:
\begin{equation}
\HP =
\left(
  \begin{array}{c|c}
             & p_1    \\
  \Q         & \vdots \\
             & p_{n-1} \\
  \hline              \\
  0 \cdots 0 & 1
  \end{array}
\right)
\label{eq:HPblocks}
\end{equation}

The \emph{absorption time} $m_i$ from vertex $i \neq s$ to the seed
vertex $s$ is the expected number of steps that a walk initiated at
$i$ will take before hitting $s$. Intuitively, as the absorption time
measures in a certain sense the proximity of vertex $i$ to vertex $s$,
vertices belonging to a good cluster $S$ for $s$, if such a cluster
exists, should have characteristically smaller absorption times to $s$
than vertices in $V \setminus S$. Note that not all graphs exhibit a
clustered structure, in which case no clustering method will be able
to pinpoint a high-quality cluster \cite{Scha07}.

It is well known that the absorption times to vertex $s = n$
can be calculated
as row sums
\begin{equation}
m_i = m_{i,1} + m_{i,2} + \ldots + m_{i, n-1}.
\label{eq:rowsums}
\end{equation}
from the {\em fundamental matrix}
\begin{equation}
\M = \I + \Q + \Q^2 + \Q^3 + \ldots =
\inv{(\I - \Q)},
\label{eq:fundamental}
\end{equation}
where $\Q$ is the matrix obtained from $\HP$ (or equivalently from $\P$)
by eliminating the row and column corresponding to vertex $s = n$
(as shown above in Equation~(\ref{eq:HPblocks})),

\begin{figure}
\centerline{\includegraphics[width=60mm]{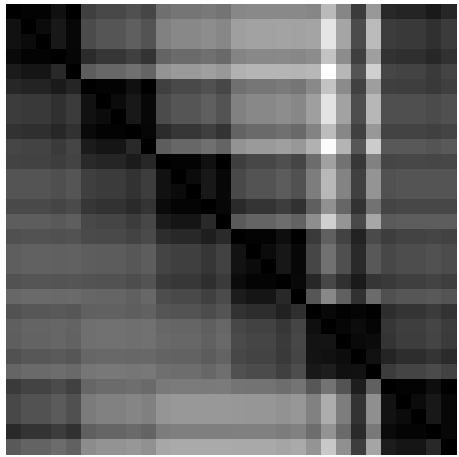}}
\caption{The absorption time matrix composed of 30 absorption-time
  vectors using each vertex of the caveman graph of Figure
  \ref{fig:caveman} in turn as a seed vertex, with white corresponding
  to the maximum $m_{i,j}$ thus obtained and black corresponding to
  the minimum $m_{i,j}$ \emph{and} the diagonal zeroes.}
\label{fig:caveman_abstime}
\end{figure}

In Figure \ref{fig:caveman_abstime}, we illustrate the absorption
times in the caveman graph of Figure~\ref{fig:caveman}: we computed
with Matlab the absorption times from all vertices to a given seed vertex
$j$, repeated the computation for each $j \in V$, and formed a matrix
where each column represents the absorption-time vector for the
corresponding vertex $j$.
The columns are ordered so that all absorption-time vectors associated
to a given cave are grouped together, before those of the next cave,
and so forth. The matrix is visualised as a gray-scale colour map by
placing a tiny \emph{black} square where either $m_{i,j} = 0$ (that is, along
the diagonal) or $m_{i,j} = 10.6$ (the minimal off-diagonal absorption
time observed), a \emph{white} square where $m_{i,j} = 319.6$
(the maximum observed), and discretising the
intermediate values to 254 gray-scale colours correspondingly. The
caves can be distinguished as dark five-by-five blocks along the diagonal,
although the matrix is somewhat too noisy to be trivially clustered.

Now consider the eigenvalue spectra of matrices $\HP$ and $\Q$.
Matrix $\HP$ is still stochastic, so it has largest eigenvalue
$\eigval{\HP}{0} = 1$, and since the chain is absorbing, all the other
eigenvalues satisfy $|\eigval{\HP}{i}| < 1$, $i = 1,\dots,n-1$.

Denote $\CQ = \sqrt{\D}\Q\invsqrt{\D}$, where $\D =
\diag(d_1,\dots,d_{n-1})$. As $\CQ$ is symmetric (it is obtained by
eliminating the last row and column from the symmetric matrix $\CP$)
and $\Q$ is similar to $\CQ$, both have a spectrum of
real eigenvalues $\spec{\Q} = \{\eigval{\Q}{1} \geq \dots \geq
\eigval{\Q}{n-1}\}$.  This spectrum is properly contained in the
interval $[-1, 1]$, because for any vertex $i \neq n$ adjacent to $n$,
$p_{in} > 0$, and so the $i$\th\ row sum of $\Q$ is less than $1$.

We claim that in fact 
\begin{equation}
\spec{\Q} = \spec{\HP} \setminus \{1\}.
\end{equation}  
To prove this claim, let namely $\lambda \neq 1$ be any non-principal
eigenvalue of $\HP$ and $\vv$ a corresponding eigenvector, so that
$\HP \vv = \lambda \vv$.  Since the $n$\th\ row of $\HP$ is zero
except for $\hat{p}_{nn} = 1$, it follows that $\lambda v_n = (\HP
\vv)_n = v_n$, and since $\lambda \neq 1$ that necessarily $v_n =
0$. Then for the $(n-1)$-dimensional vector $\vv' =
(v_1,\dots,v_{n-1})$ and for any $i = 1,\dots,n-1$ it holds that:
\begin{equation}
\begin{array}{rcl}
\displaystyle
(\Q \vv')_i
& = & \displaystyle
      \sum_{j=1}^{n-1} p_{ij}v'_j
  =   \sum_{j=1}^{n-1} p_{ij}v_j
  =   \sum_{j=1}^n p_{ij}v_j - p_{in}v_n \\
& = & (\HP \vv)_i - v_n p_{in}
  =   (\HP \vv)_i \\
& = & \lambda v_i
  =   \lambda v'_i.
\end{array}
\label{eq:Qeigvect}
\end{equation}
Consequently, $\vv'$ is an eigenvector associated to
eigenvalue $\lambda$ of $\Q$. Since $\lambda$ was chosen
arbitrarily from $\spec{\HP}\setminus\{1\}$, this
establishes that $\spec{\HP}\setminus\{1\} \subseteq \spec{\Q}$.
For the converse direction, a similar argument shows that if
$\vv' =  (v_1,\dots,v_{n-1})$ is an eigenvector associated
to an eigenvalue $\lambda$ of $\Q$, then the vector
$\vv =  (v_1,\dots,v_{n-1},0)$ is an
eigenvector associated to eigenvalue $\lambda$ of $\HP$.

\section{Spectral methods for bipartitioning}

\subsection{Fiedler vectors}

Spectral clustering of points in space, often modelled as (complete)
weighted graphs, is a widely studied topic \cite{HKK07, KVV04}. In the
context of graphs, the technique is usually applied so that some right
eigenvector associated to the smallest nonzero eigenvalue
$\leigval{\L}{1}$ of $\L$ is used to produce a bipartitioning of the
graph such that those vertices that have negative values in the
eigenvector form one side of the bipartition $S$ and the vertices with
positive values are the other side $S \setminus V$.  These
eigenvectors are called \emph{Fiedler vectors} following
\cite{Fied73,Fied75}, where the technique was first proposed.  The
corresponding eigenvectors based on $\CL$ are called \emph{normalised}
Fiedler vectors. The works on Fiedler-vector based spectral clustering
are numerous and go back for decades \cite{SpTe96,QiHa06,HoSu99}

\begin{figure}
\centerline{\includegraphics[width=128mm]{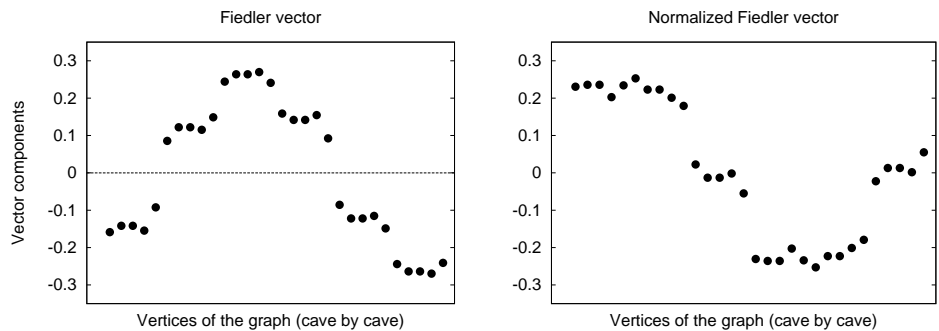}}
\caption{The components of the Fiedler vector (left) and the
  normalised Fiedler vector (right) for the caveman graph of Figure
  \ref{fig:caveman}. For the human eye, the six-cluster structure is
  evident in the Fiedler vectors, whereas in the normalised Fiedler vector
  the vertices are grouped into four clusters (two of them consisting
  of two caves).}
\label{fig:laplace}
\end{figure}

For our example graph illustrated in Figure \ref{fig:caveman}, such a
bipartition based on $\L$ puts three of the caves in $S$ such that it
assigns negative values to every other cave along the cycle of six
caves. Using the eigenvector of $\CL$, however, assigns only negative
values in the vector and does not yield an intuitive division that
preserves the caves. The two vectors are visualised in Figure
\ref{fig:laplace}.

\begin{figure}
\centerline{\includegraphics[width=128mm]{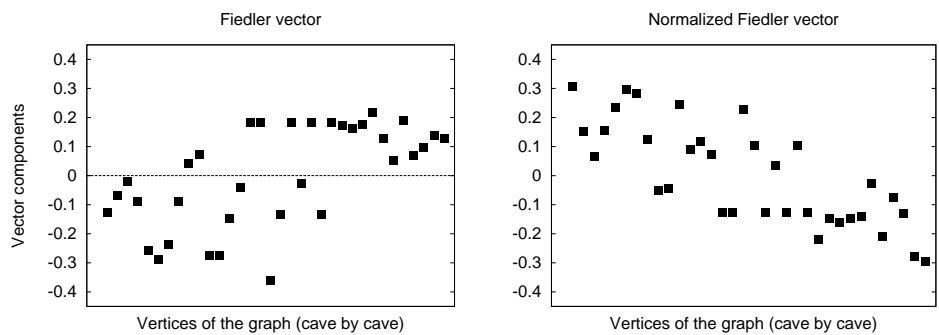}}
\caption{The components of the Fiedler vector (left) and the
  normalised Fiedler vector (right) for the karate club graph of Figure
  \ref{fig:karate}. The vertices can be classified in two groups:
  those with positive values in the Fiedler vector and those with
  negative values.}
\label{fig:lapkarate}
\end{figure}

If there are only two natural clusters in the graph, such bipartition
works nicely. An example is the Zachary karate club network of Figure
\ref{fig:karate}: the corresponding Fiedler vectors are shown in
Figure \ref{fig:lapkarate}. Also, recursively performing bipartitions
on the subgraphs induced by $S$ and $V \setminus S$ will help cluster
the input graph $G$ in more than two clusters, but a \emph{stopping
  condition} needs to be imposed to determine when to stop
bipartitioning the resulting subgraphs further.

\subsection{Spectral partitioning as integer program relaxation}

The use of Fiedler vectors for graph bipartitioning can be motivated
as follows (see for example \cite{HKK07}). Denote a \emph{cut}
(bipartition) of a graph $G = (V,E)$ into vertex sets $S$ and $\bar{S}
= V \setminus S$ as $\Scut$.  The {\em capacity} of a cut $\Scut$ is
defined as
\begin{equation}
C\Scut = \left|\{\{i,j\} \in E : i \in S, j \in \Sbar\}\right|.
\end{equation}
A cut $\Scut$ can be conveniently represented by an indicator vector
$\vv \in \{+1,-1\}^n$, where $v_i = +1$ if $i \in S$, and $v_i = -1$
if $i \in \Sbar$. 

Then
\begin{equation}
  C\Scut = \frac{1}{4} \sum_{i \sim j} (v_i - v_j)^2,
\end{equation}
where the sum is over all the (undirected) edges $\{i,j\} \in E$.

For simplicity, assume now that $|V| = n$ is even, and
consider the task of finding an {\em optimal bisection} of $G$,
i.e.\ a cut $\Scut$ that satisfies $|S| = |\Sbar| = n/2$
and minimises $C\Scut$ subject to this condition.

This is equivalent to finding an indicator vector
$\vv \in \{+1,-1\}^n$ that satisfies $\sum_i v_i = 0$ and
minimises the quadratic form $\sum_{i \sim j} (v_i - v_j)^2$,
or equivalently (since $n$ is fixed) minimises the ratio:
\begin{align*}
  \frac{\frac{1}{4}\sum_{i \sim j} (v_i - v_j)^2}{n/4}
  & = \frac{\sum_{i \sim j} (v_i - v_j)^2}{n}\\
  & = \frac{\sum_{i \sim j} (v_i - v_j)^2}{\sum_i v_i^2}.
\end{align*}
Since the all-ones vector $\one$ is associated to the
eigenvalue $\leigval{\L}{0} = 0$, we have by the
Courant-Fischer characterisation of the smallest
nonzero eigenvalue $\leigval{\L}{1}$:
\begin{equation}
\leigval{\L}{1}
  \quad = \quad
  \min_{\vv \bot \one}\frac{\transpose{\vv} L \vv}{\transpose{\vv} \vv}
  \quad = \quad
  \min_{\sum_i v_i = 0}
    \frac{\sum_{i \sim j} (v_i - v_j)^2}
         {\sum_i v_i^2},
\end{equation}
where the minimum is taken over all vectors $\vv \neq 0$
satisfying the given condition. Since we can without
loss of generality also constrain the minimisation to,
say, the vectors of norm $\|\vv\|^2 = n$, we see that
the task of finding a Fiedler vector of $G$ is in fact
a fractional relaxation of the combinatorial problem of
determining an optimal bisection of $G$.

This correspondence motivates the previously indicated 
spectral approach to bisectioning a connected graph $G$
\cite{DoHo73,Fied73}:
\begin{enumerate}
\item Compute Fiedler vector $\vv \in \RR^n$ of $G$.
\item Determine cut $\Scut$ by rule:
  \begin{equation}\left\{
    \begin{array}{lcl}
      v_i > \theta & \Rightarrow & i \in S, \\
      v_i < \theta & \Rightarrow & i \in \Sbar,
    \end{array}
    \right.
    \end{equation}
\end{enumerate}
where $\theta$ is the median value of the $v_i$'s.

The use of \emph{normalised} Fiedler vectors to graph
bipartitioning was explored in \cite{ShMa00},
where it was shown that Fiedler vectors of $\CL$
yield fractionally optimal graph bipartitions
according to the {\em normalised cut capacity} measure:
\begin{equation}\Chat\Scut = \frac{C\Scut}{\Vol(S)} +
               \frac{C\Scut}{\Vol(\Sbar)},
\end{equation}
where $\Vol(S) = \sum_{i \in S} d_i$.

\ignore{
Note that this is quite close to the notion of {\em conductance}
of cut $\Sbar$:
\begin{equation}
\Phi\Scut = \max\{\frac{C\Scut}{\Vol(S)},
                  \frac{C\Scut}{\Vol(\Sbar)}\}.
\end{equation}
}

Since $\u$ is an eigenvector of $\CL$ with
eigenvalue $\lambda$ if and only if $\vv = \invsqrt{\D} \u$
is eigenvector of $\D^{-1}\L$ with eigenvalue $\lambda$,
the eigenvalue $\leigval{\CL}{1}$ can be characterised
in terms of a ``degree-adjusted'' Rayleigh quotient:
\begin{equation}
\leigval{\CL}{1}
  = \min_{\u \bot \sqrt{\D}\one} 
\frac{\transpose{\u} \CL \u}{\transpose{\u} \u}
  = \min_{\vv \bot \D\one}
    \frac{\sum_{i \sim j} (v_i - v_j)^2}{\sum_i d_i v_i^2}.
\end{equation}
\ignore{
Note that
\begin{equation}\vv \bot D\one \iff \sum_i v_i d_i = 0.\end{equation}
}

Since $\u$ is an eigenvector of $\CL$ with eigenvalue $\lambda$ if and
only if $\vv = \invsqrt{\D} \u$ is eigenvector of $\D^{-1}\L$ with
eigenvalue $\lambda$, the eigenvalue $\leigval{\CL}{1}$ can be
characterised in terms of a ``degree-adjusted'' Rayleigh quotient:
\begin{equation}
\leigval{\CL}{1}
  = \min_{\u \bot \sqrt{\D}\one} 
\frac{\transpose{\u} \CL \u}{\transpose{\u} \u}
  = \min_{\vv \bot \D\one}
    \frac{\sum_{i \sim j} (v_i - v_j)^2}{\sum_i d_i v_i^2}.
\end{equation}

A natural extension of the spectral clustering idea to the local
clustering context is to consider the Laplacian $\L$ or $\CL$ together
with the \emph{Dirichlet boundary condition} that only clustering
vectors $\vv$ with the seed vertex $v_s$ fixed to some particular
value are acceptable solutions.

We follow \cite{Chun97,ChEl02} in using the normalised Laplacian $\CL$
and choosing $v_s = 0$, or equivalently $u_s = (\sqrt{\D}\vv)_s = 0$
as the boundary condition.  We thus aim to cluster according to the
``Dirichlet-Fiedler vector'' minimising the constrained Rayleigh
quotient:
\begin{equation}
\min_{\u: u_s = 0} \frac{\transpose{\u} \CL \u}{\transpose{\u} \u}
  = \min_{\vv: v_s = 0}
    \frac{\sum_{i \sim j} (v_i - v_j)^2}{\sum_i d_i v_i^2}.
\label{eq:rayleigh}
\end{equation}

For notational simplicity, assume again that $s = n$, and observe
that for every vector $\u = (u_1,\dots,u_{n-1},0)$, the value of
the Rayleigh quotient in equation~(\ref{eq:rayleigh}) is the same
as the value of the $(n-1)$-dimensional quotient with respect to vector
$\u' = (u_1,\dots,u_{n-1})$ and Laplacian $\CL'$ which equals
$\CL$ with its $n$\th\ row and column removed. Thus, our clustering
vector $\vv$ is, except for the final zero, the one minimising: 
\begin{equation}
\min_{\u'} \frac{\transpose{(\u')} \CL' \u'}{\transpose{(\u')} \u'}
  = \min_{\vv'}
    \frac{\sum_{i \sim j} (v'_i - v'_j)^2}{\sum_i d_i (v'_i)^2},
\label{eq:rayleighp}
\end{equation}
i.e.\ $\vv' = \invsqrt{\D}\u'$ for the principal eigenvector 
$\u'$ of the Laplacian $\CL'$. Let us denote $\vv = \vf$ and call
this the \emph{local Fiedler vector} associated to graph $G$ and
seed vertex $s = n$. 

\section{Local Fiedler vectors and absorption times of random walks}

We shall now show that the components of the local Fiedler vector $\vf
= (v_1, \dots, v_{n-1})$ are in fact approximately proportional to the
absorption times $m_i$ discussed in Section~\ref{sec:random walks}.
The connection between the absorption time provides a natural
interpretation to the notion of the local Fiedler vector, and yields
further support to the idea of local clustering by constrained
spectral techniques. Previously random walks and spectral clustering
have been jointly addressed by Meila and Shi \cite{MiSh01} and local
clustering by PageRank by Andersen, Chung, and Lang \cite{dirlocal}.
Important papers linking structural properties of graphs to convergence
rates of random walks via spectral techniques are \cite{Alon86,SiJe89}.

Observe first, from equation~(\ref{eq:lapprob}), that:
\begin{equation}
\CL'  =   \I - \sqrt{\D}\Q\invsqrt{\D} = \I - \CQ,
\end{equation}
where $\Q$ is as in Equation~(\ref{eq:HPblocks})
and $\D = \diag(d_1,\dots,d_{n-1})$.

Since $\CQ$ is similar to $\Q$, its spectrum satisfies:
\begin{equation}
\spec{\CQ} = \spec{\Q} = \spec{\HP} \setminus \{1\}.
\end{equation}
Thus, $\leig \neq 0$ is an eigenvalue of $\CL'$
if and only if $\eig = 1-\leig \neq 1$
is an eigenvalue of both $\CQ$ and $\Q$.
Moreover, if $\u$ is an eigenvector associated to eigenvalue
$\eig$ in $\CQ$, then $\vv = \invsqrt{\D} \u$ is an
eigenvector associated to the same eigenvalue in $\Q$.

Let then the eigenvalues of $\CQ$ (or equivalently $\Q$) be $1 >
\eig_1 \geq \dots \geq \eig_{n-1} \geq 0$.  Since $\CQ$ is symmetric, it
has a corresponding orthonormal system of eigenvectors
$\u_1,\dots,\u_{n-1}$ and a representation:
\begin{equation}
  \CQ = \displaystyle\sum_{i = 1}^{n - 1} \eig_i \u_i \transpose{\u_i}.
\label{eq:cqsumrep}
\end{equation}
Denoting the component matrices
$\U_i = \u_i \transpose{\u_i}$, we observe that by
orthogonality of the eigenvectors we have
$\U_i \U_j = 0$ for $i \neq j$, and by normality $\U_i^2 = \U_i$.
From these two observations it follows that:
\begin{equation}
\CQ^t = \displaystyle\sum_{i = 1}^{n-1} \eig_i^t \U_i,
  \qquad \text{for } t = 0,1,\dots
\end{equation}
Since $\Q = \invsqrt{\D} \CQ \sqrt{\D}$,
we obtain from this for $\Q^t$ the representation:
\begin{equation}
\Q^t =\invsqrt{\D} \CQ^t \sqrt{\D}
  = \displaystyle
  \sum_{i = 1}^{n - 1} \eig_i^t
    (\invsqrt{\D} \u_i)(\transpose{\u_i} \sqrt{\D})
  = \displaystyle
  \sum_{i = 1}^{n-1} \eig_i^t \vv_i \transpose{\vv_i} \D,
\end{equation}
where $\vv_i = \invsqrt{\D}\u_i$ is an eigenvector associated
to eigenvalue $\lambda_i$ in $\Q$.

Substituting this to Equation~(\ref{eq:fundamental}) and denoting
the $(n-1)$-dimensional all-ones vector by $\one$, we thus obtain
an expression for the vector $\m$ of absorption times $m_i$
in terms of the eigenvalues and eigenvectors of $\Q$, or
equivalently $\CQ$:
\begin{equation}
\begin{array}{rcl}
\m &=& \displaystyle
  \sum_{t = 0}^\infty \Q^t \one \\
  &=& \displaystyle
  \sum_{t = 0}^\infty
    \sum_{i = 1}^{n-1} \eig_i^t \vv_i \transpose{\vv_i} \D \one \\
  &=& \displaystyle
  \sum_{t = 0}^\infty
    \left(\sum_{i = 1}^{n-1} \eig_i^t \vv_i \transpose{\vv_i} \right) \d ,
\end{array}
\label{eq:abssum}
\end{equation}
where $\d = \transpose{(d_1,\dots,d_{n-1})}$.

Now if the principal eigenvalue $\eig_1$ is well-separated
from the others, i.e.\ if the ratio $|\eig_i / \eig_1|$ is small
for $i > 1$, this yields a good approximation for $\m$:
\begin{equation}
\begin{array}{rcl}
\m &=& \displaystyle
 \one +
 \sum_{t = 1}^\infty
   \eig_1^t \left(\vv_1 \transpose{\vv_1} \d +
   \underbrace{
     \sum_{i = 2}^{n-1}
       \left(\frac{\eig_i}{\eig_1}\right)^t \vv_i \transpose{\vv_i} \d
   }_{\text{small-norm "noise"}}
   \right) \\
  &\approx& \displaystyle
  \one + \sum_{t = 1}^\infty \eig_1^t \vv_1 \transpose{\vv_1} \d \\
  &=& \displaystyle
  \one + \frac{\eig_1}{1 - \eig_1} \vv_1 \transpose{\vv_1} \d.  
\end{array}
\label{eq:approxabsvect}
\end{equation}
Even in cases where there is no evident gap in the spectrum and hence
near-equality cannot be assumed, we have found in our experiments that
the approximations obtained are near-perfectly correlated with the exact
absorption times for a variety of graphs.

We study three example graphs to point out the strengths and
weaknesses of the proposed approximation. The first example graph is
the clustered but highly symmetric caveman graph of Figure
\ref{fig:caveman}, where the symmetries present cause problems for the
proposed approximation.  Our second example is the karate club network
shown in Figure \ref{fig:karate}. The third example graph is a uniform
random graph $\mathcal{G}_{n, p}$, with $n = 100$ and $p = 0.1$
\cite{Gilb59}, which by definition has no clear cluster structure, and
hence the absorption times cannot be expected to have interesting patterns.

\begin{figure}
\centerline{\includegraphics[width=128mm]{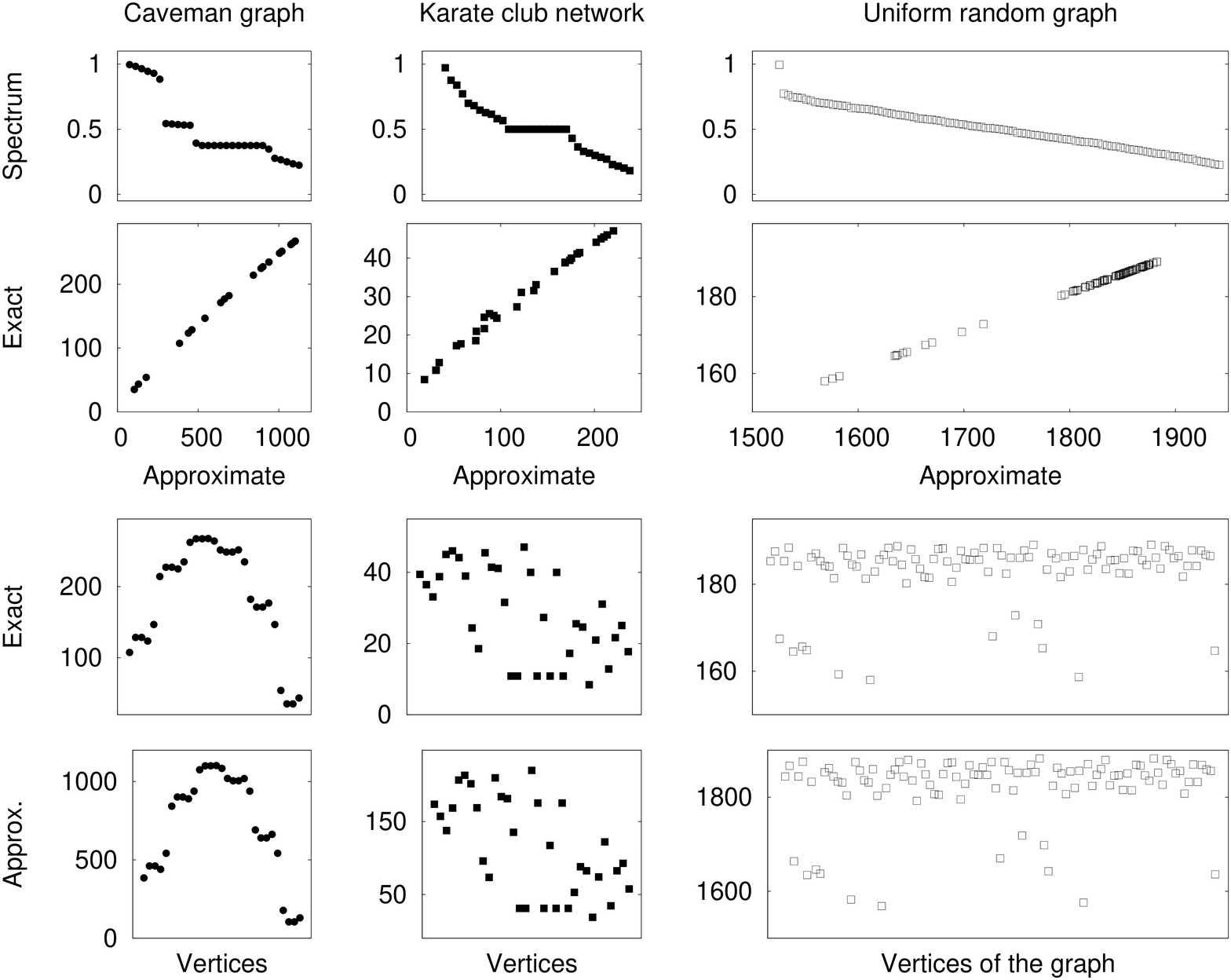}}
\caption{Comparisons of approximate and exact spectral computations
  for three
  example graphs: the small graphs of Figures \ref{fig:caveman} and
  \ref{fig:karate}, and a uniform random graph $\mathcal{G}_{n, p}$,
  using a random vertex as the seed vertex.
  The top row presents the sorted spectra of the
  $\HP$ matrices of the graphs,
  the second row plots the approximate and exact absorption-time
  values for the given seed vertex against each other,
  and the lowest two rows indicate the exact and approximate
  absorption-time values as ordered by vertex number.
  The bottom rows can be seen as illustrating the quantities'
  capability of distinguishing the cluster of the seed vertex
  (low values) from the other vertices (high values).}
\label{fig:examples}
\end{figure}

In Figure~\ref{fig:examples}, we show comparisons of some approximate
and exact spectral computations for three example graphs.
In each case, the highest-numbered vertex
of the graph has been chosen as the unique seed vertex.
It can be noted, from the top row of plots in Figure~\ref{fig:examples},
that the spectra of the graphs' $\HP$ matrices do not exhibit
large gaps between their second and third largest eigenvalues.
Thus, it can not be expected \emph{a priori} that the Fiedler-vector
based approximations to the absorption times, from
Equation~(\ref{eq:approxabsvect}), would be even
of the same magnitude as the exact ones, as calculated from
Equations~(\ref{eq:rowsums}) and~(\ref{eq:fundamental}).
(Observe also how the structure of the caveman graph is reflected
in the corresponding $\HP$ spectrum: a notable eigenvalue gap
occurs after the six largest eigenvalues, each representing the
dominant convergence behaviour of one of the clusters.)

Correlations between the approximate and exact absorption times
are apparent in the quantile-quantile plots presented in the
second row of Figure~\ref{fig:examples}: here the values
group diagonally when a linear dependency exists.
The correlation is very high in all
cases: $0.99863$ for the caveman graph, $0.99636$ for the karate club
network, and $0.99999$ for the uniform random graph.

The two lowest rows in Figure~\ref{fig:examples} present the
actual values of the exact and approximate absorption-time vectors,
indexed by vertex number. These plots illustrate the usefulness
of these quantities for performing a two-classification of
the vertices into the local cluster of the seed vertex
(low values) versus the other vertices (high values).
In fact, for the caveman graph, the full
six-cluster structure is visible. In the karate club network it can be
seen that two groups are present: one with high values and another one
with low values. (Cf.\ Figure \ref{fig:karateclass}, which indicates
the ``ground truth'' clustering of the vertices in this graph.)
As expected, the uniform random graph
reveals no significant cluster structure, but the vertices near the
seed vertex can be identified by their lower values, whereas most of
the graph has another, higher value. 

In practice, it is not always interesting to compute the absorption
times for all vertices, especially in local computation, in which case
we may only have approximated some of the components of the Fiedler
vector. For these situations, we may write the $k$\th\ component of
the result vector as
\begin{equation}
\begin{array}{rcl}
(\Q^t \one)_k &=& \displaystyle
  (\sum_{i = 1}^{n-1} \eig_i^t \vv_i \transpose{\vv_i} \D \one)_k \\
  &=& \displaystyle
  (\sum_{i = 1}^{n-1} \eig_i^t (\transpose{\vv_i} \d) \vv_i)_k \\
  &=& \displaystyle
  \sum_{i = 1}^{n-1} \eig_i^t (\vv_i)_k 
  \underbrace{
    \sum_{\ell = 1}^{n-1} (\vv_i)_\ell (\d)_{\ell}
  }_{\constant_i}.
\end{array}
\end{equation}
From this we obtain for the absorption time from vertex $k$
to vertex $s$ the expression
\begin{equation}
\begin{array}{rcl}
m_k &=& \displaystyle
  \sum_{t = 0}^\infty (\Q^t \one)_k \\
  &=& \displaystyle
  1 + \sum_{t = 1}^\infty \eig_1^t
      \left(\constant_i \cdot (\vv_i)_k 
        + \sum_{i = 2}^{n-1} \left(\frac{\eig_i}{\eig_1}\right)^t
            \constant_i \cdot (\vv_i)_k
      \right) \\
  &\approx& \displaystyle
  1 + \sum_{t = 1}^\infty \eig_1 \cdot \constant_1 \cdot (\vv_1)_k \\
  & = & \displaystyle
  1 + \underbrace{
        \frac{\eig_1}{1 - \eig_1} \cdot \constant_1
      }_{\constant'} \cdot (\vv_1)_k.
\end{array}
\label{eq:fiedlerabsapprox}
\end{equation}
Now for a given graph $G$, $\constant'$ is a constant and so we obtain
the very simple approximate correspondence $\m \approx \one +
\constant' \vf$ between the absorption time vector $\m$ and the local
Fiedler vector $\vf = \vv_1$.

\begin{figure}
\centerline{\includegraphics[width=128mm]{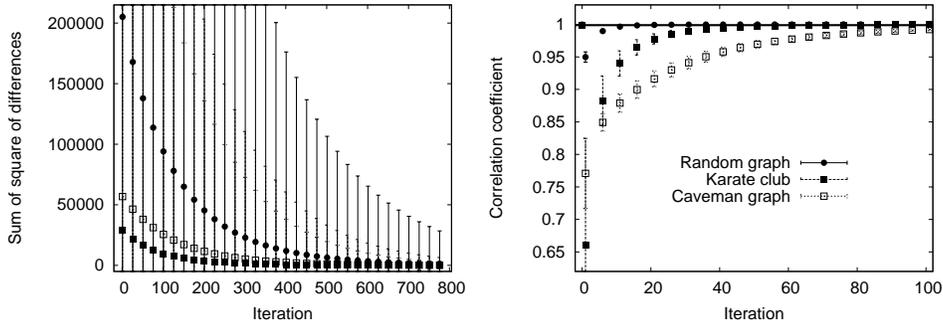}}
\caption{The sum of squares of the difference from the exact
  absorption-time (on the left) of estimate vectors with different
  cutoff values for approximating through Equation~(\ref{eq:abssum}) and
  Pearson correlation between the exact and the estimate vectors (on
  the right) for the three example graphs: the small graphs of Figures
  \ref{fig:caveman} and \ref{fig:karate}, and the $\mathcal{G}_{n,
    p}$. The values shown are averaged over the vertex sets of the two
  small graphs and over a set of 30 vertices selected uniformly at
  random for the $\mathcal{G}_{n, p}$ graph. The smallest standard
  deviation corresponds to the caveman graph and the largest to the
  uniform random graph. The horizontal lines (all three overlap
  between $0.980$ and $0.997$) correspond to the average correlation
  coefficients between the exact and the approximate absorption times
  of Equation~(\ref{eq:approxabsvect}).}
\label{fig:absconv}
\end{figure}

In order to compare the quality of the approximation as well as to
illustrate the computational load in approximating by summing term by
term the series of Equation~(\ref{eq:abssum}), we calculated for each
cutoff length the sum of squares of the differences between the
partial sums and the exact absorption times, divided by the order of
each of the three example graphs: the graph of Figure
\ref{fig:caveman}, the Zachary karate club graph of Figure
\ref{fig:karate}, and the uniform random graph $\mathcal{G}_{n,
  p}$. The resulting values over the set of vertices are shown in
Figure \ref{fig:absconv} (on the left) together with the Pearson
correlations (on the right) achieved at each iteration.  In both
plots, mean and standard deviation are shown.

\section{Local approximation of Fiedler vectors}

We take as a starting point the Rayleigh quotient of
Equation~(\ref{eq:rayleighp}). Since we are free to normalise our
eventual Fiedler vector $\vf$ to any length we wish, we can constrain
the minimisation to vectors $\vv$ that satisfy, say, $\|\vv\|_2^2 = n
= |V|$.  Thus, the task becomes one of finding a vector $\vv$ that
satisfies for a given $s \in V$:
\begin{equation}   
  \vf \quad = \quad
  \argmin \bigg \{\sum_{j \sim k} (v_j - v_k)^2 :
  v_s = 0,\; \|\vv\|_2^2 = n \bigg\}. 
\label{eq:fiedler}
\end{equation}
We can solve this task approximately by reformulating the requirement
that $\|\vv\|_2^2 = n$ as a ``soft constraint'' with weight $c > 0$,
and minimising the objective function
\begin{equation}   
  f(\vv) \quad = \quad \frac{1}{2} \sum_{j \sim k}
  \bigg(v_j - v_k\bigg)^2 +
  \frac{c}{2} \cdot
  \bigg(n - \sum_j v_j^2\bigg)
\label{eq:soft_fiedler}
\end{equation}
by gradient descent. Since the partial derivatives of $f$ have
the simple form
\begin{equation}   \label{eq:soft_partials}
\frac{\partial f}{\partial v_j} \quad = \quad
   - \sum_{k \sim j} v_k+ (\deg{j} - c) \cdot v_j,
\end{equation}
the descent step can be computed locally at each vertex at time $t +
1$, based on information about the values of the vector $\vv$ at time
$t$, denoted by $\tilde{\vv}(t)$, for the vertex itself and its
neighbours:
\begin{equation} \label{eq:grad_desc} \tilde{v}_j(t+1) \quad = \quad
  \tilde{v}_j(t) + \delta \cdot \left(\sum_{k \sim j} \tilde{v}_k -
  (\deg{j} - c) \cdot \tilde{v}_j\right),
\end{equation}
where $\delta > 0$ is a parameter determining the speed of the descent.

Assuming that the natural cluster of vertex $s$ is small compared to
the order of the graph $n$, the normalisation $\|\vv\|_2^2 = n$ entails
that most vertices $j$ in the network will have $v_j \approx 1$.  Thus
the descent iterations~(\ref{eq:grad_desc}) can be started from an
initial vector $\tilde{\vv}(0)$ that has $\tilde{v}_s(0) = 0$ for the
seed vertex $s \in V$ and $\tilde{v}_k(0) = 1$ for all $k \neq i$.
The estimates need then to be updated at time $t > 0$ only for those
vertices $j$ that have at least one neighbour $k$ such that
$\tilde{v}_k(t-1) < 1$.

Balancing the constraint weight $c$ against the speed of gradient
descent $\delta$ naturally requires some care.  We have obtained
reasonably stable results with the following heuristic: given an
estimate $\bar{k}$ for the average degree of the vertices in the
network, set $c = 1/\bar{k}$ and $\delta = c/10$. The gradient
iterations (\ref{eq:grad_desc}) are then continued until all the
changes in the $v$-estimates are below $\varepsilon = \delta/10$.  We
leave the calibration of these parameters to future work.

The (approximate) Fiedler values thus obtained represent
proximity-values of the vertices in $V$ to the cluster of vertex $s$.
Determining a bisection into $S$ and $V \setminus S$ is now a
one-dimensional two-classification task that can in principle be
solved using any of the standard pattern classifiers, such as
variations of the basic $k$-means algorithm~\cite{HaWo79}.

\begin{figure}
\centerline{\includegraphics[width=128mm]{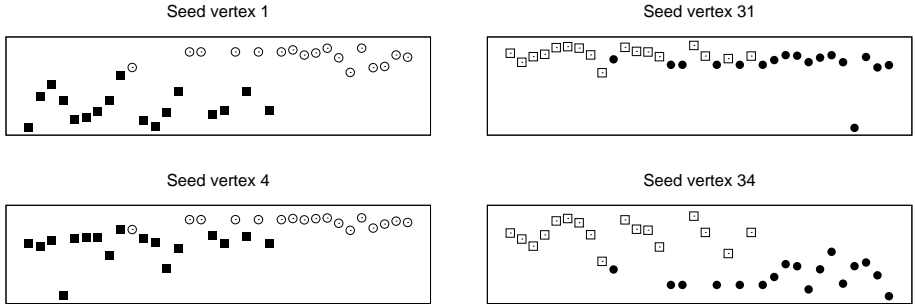}}
\caption{Four examples of two-classifying vertices of the Zachary
  karate club graph. The examples on the left have the seed vertex
  among the ``rectangles'' of Figure \ref{fig:karate} and the examples
  of the right have the seed vertex among the ``circles''. The
  vertices are ordered by their label in Figure \ref{fig:karate} and a
  zero has been inserted to represent the absorption time to the seed
  vertex itself. The group in which the seed belongs is drawn in black
  and the other group in white.}
\label{fig:karateclass}
\end{figure}

We illustrate the applicability approximate absorption times for
clustering the karate club network (Figure \ref{fig:karate}).  The
approximate absorption times shown in Figure \ref{fig:karateclass} are
computed directly with Equation~(\ref{eq:fiedlerabsapprox}): the group
structure is seen to be strong when the seed vertex is one of the
central members of the group, whereas the classification task is
harder for the ``border'' vertices, as can be expected. For more
extensive examples of clustering with the locally computed
approximates, we refer the reader to previous work \cite{OrSc05}.

\section{Conclusions and further work}

In this work we have derived an expression for the absorption times to
a single absorbing vertex $s$ in a simple random walk in an
undirected, unweighted graph in terms of the spectrum of the
normalised Laplacian matrix of the graph. We have shown that by
only knowing the Fiedler vector corresponding to $s$ on the boundary and
the corresponding eigenvalue provides an approximation of the
absorption times if the spectrum of the graph presents a gap after the
first eigenvalue. Experimentally we have confirmed that the values given
by the approximation are nearly perfectly correlated with the exact
absorption times even in the absence of such a gap. 

Our motivation is to use the absorption times into a seed vertex $s$
as a measure of proximity in two-classifying the graph into two
partitions: vertices that are ``relevant'' to the seed vertex and
other vertices. Hence, not knowing the exact values but rather another
vector of perfectly correlated values is sufficient for separating
between the vertices with higher values from those with lower values
(which is the classical two-classification task).

Such a two-partition of a graph is known as local clustering. In order
for the proposed values to be locally computable, we have also presented
a gradient-descent method to approximate the Fiedler vector using only
local information in the graph. The method iteratively processes the
neighbourhoods of vertices starting from the seed vertex and expanding
outwards within the group of potentially ``relevant'' vertices,
without any need to process other parts of the graph. We have
illustrated the potential of these vectors in two-classification for local
clustering on a classical example graph representing a social network.

In further work, we seek to study further the effects of the presence
or absence of a spectral gap in the input graph into the approximation
proposed. We also want to calibrate the parameters of the locally
computable approximation in such a way that no a priori knowledge of
the input graph would be needed, but that the method would rather
adapt to the structure of the graph at runtime by dynamic parameter
adjustment. Of additional interest are extensions of this work to
weighted and directed graphs as well as case studies of applications
of local clustering. We also contemplate possible uses for approximate
absorption times in resolving other problems of interest that involve
complex systems represented as graphs.

\section*{Acknowledgements}

The work of Orponen and Schaeffer was supported by the Academy of
Finland under grant 206235 (ANNE, 2004--2006). Schaeffer and Avalos
received support from the UANL under grant CA1475-07 and from PROMEP
under grant 103,5/07/2523. Avalos also thanks CONACYT for support.

A preliminary report on parts of this work was presented as ``Local
clustering of large graphs by approximate Fiedler vectors'' by
P. Orponen and S. E. Schaeffer, at the Fourth International Workshop
on Efficient and Experimental Algorithms in Santorini, Greece, May
2005. The current work was presented at The Fifteenth Conference of the
International Linear Algebra Society (ILAS) in Canc\'{u}n, Quintana
Roo, Mexico, in June 2008.

%\bibliography{OrScAv2008} 
%\bibliographystyle{unsrt}

\end{document}